\documentstyle[twoside,fleqn,espcrc2]{article}

\newcommand{\AmS}{{\protect\the\textfont2
  A\kern-.1667em\lower.5ex\hbox{M}\kern-.125emS}}

\title{
Precanonical perspective 
in quantum gravity\thanks{
Based on the talk at the Third Meeting on Constrained Dynamics 
and Quantum Gravity QG99 
(Villasimius, Sardinia, Italy, Sept. 13-17, 1999). 
To appear in {\em Nucl. Phys. B Proc. Suppl.} (2000) }
}

\author{
I. V. Kanatchikov\address{
 Laboratory of Analytical Mechanics and Field Theory, 
 Institute of Fundamental Technological Research, 
 Polish Academy of Sciences, Warsaw 00-049, Poland 
 }%
{$^,$}\address{Institute of Theoretical Physics, 
Friedrich Schiller University, 
Jena 07743, Germany 
}%
{$^,$}\thanks{On leave from Tallinn Technical University, 
Tallinn, Estonia }  
}

\begin{document}

\begin{abstract}
Quantization of general relativity in metric variables 
using ``precanonical'' quantization based on  
the De Donder-Weyl covariant Hamiltonian formulation 
is outlined. Elements of classical geometry needed to 
formulate the 
wave equation emerge from a self-consistency with the 
underlying quantum dynamics of the metric  
in this sense ensuring  
the background independence of the formulation.  

\end{abstract}

\maketitle 

 \vspace*{-94mm}\vspace*{-2mm} 
\hbox to 6.25truein{
\footnotesize\it 
\hfil \hbox to 0 truecm{\hss 
\normalsize\rm 
{\sf  FSU TPI \, 13/99} { }}\vspace*{-3.5mm}}
\hbox to 6.2truein{
\vspace*{-1mm}\footnotesize 
\hfil 
} 
\hbox to 6.25truein{
  \footnotesize 
\hfil \hbox to 0 truecm{ 
\hss \normalsize  
\sf gr-qc/0004066 
}  
}

  
\vspace*{62mm} \vspace*{16mm} 



\newcommand{\beq}{\begin{equation}}
\newcommand{\eeq}{\end{equation}}
\newcommand{\beqa}{\begin{eqnarray}}
\newcommand{\eeqa}{\end{eqnarray}}
\newcommand{\nn}{\nonumber}

\newcommand{\half}{\frac{1}{2}}

\newcommand{\xt}{\tilde{X}}

\newcommand{\uind}[2]{^{#1_1 \, ... \, #1_{#2}} }
\newcommand{\lind}[2]{_{#1_1 \, ... \, #1_{#2}} }
\newcommand{\com}[2]{[#1,#2]_{-}} 
\newcommand{\acom}[2]{[#1,#2]_{+}} 
\newcommand{\compm}[2]{[#1,#2]_{\pm}}

\newcommand{\lie}[1]{\pounds_{#1}}
\newcommand{\co}{\circ}
\newcommand{\sgn}[1]{(-1)^{#1}}
\newcommand{\lbr}[2]{ [ \hspace*{-1.5pt} [ #1 , #2 ] \hspace*{-1.5pt} ] }
\newcommand{\lbrpm}[2]{ [ \hspace*{-1.5pt} [ #1 , #2 ] \hspace*{-1.5pt}
 ]_{\pm} }
\newcommand{\lbrp}[2]{ [ \hspace*{-1.5pt} [ #1 , #2 ] \hspace*{-1.5pt} ]_+ }
\newcommand{\lbrm}[2]{ [ \hspace*{-1.5pt} [ #1 , #2 ] \hspace*{-1.5pt} ]_- }
\newcommand{\pbr}[2]{ \{ \hspace*{-2.2pt} [ #1 , #2 ] \hspace*{-2.55pt} \} }
\newcommand{\we}{\wedge}
\newcommand{\dv}{d^V}
\newcommand{\nbrpq}[2]{\nbr{\xxi{#1}{1}}{\xxi{#2}{2}}}
\newcommand{\lieni}[2]{$\pounds$${}_{\stackrel{#1}{X}_{#2}}$  }

\newcommand{\rbox}[2]{\raisebox{#1}{#2}}
\newcommand{\xx}[1]{\raisebox{1pt}{$\stackrel{#1}{X}$}}
\newcommand{\xxi}[2]{\raisebox{1pt}{$\stackrel{#1}{X}$$_{#2}$}}
\newcommand{\ff}[1]{\raisebox{1pt}{$\stackrel{#1}{F}$}}
\newcommand{\dd}[1]{\raisebox{1pt}{$\stackrel{#1}{D}$}}
\newcommand{\nbr}[2]{{\bf[}#1 , #2{\bf ]}}
\newcommand{\der}{\partial}
\newcommand{\oo}{$\Omega$}
\newcommand{\Om}{\Omega}
\newcommand{\om}{\omega}
\newcommand{\eps}{\epsilon}
\newcommand{\si}{\sigma}
\newcommand{\Lm}{\bigwedge^*}

\newcommand{\inn}{\hspace*{2pt}\raisebox{-1pt}{\rule{6pt}{.3pt}\hspace*
{0pt}\rule{.3pt}{8pt}\hspace*{3pt}}}
\newcommand{\sro}{Schr\"{o}dinger\ }
\newcommand{\bm}{\boldmath}
\newcommand{\vol}{\omega}
               \newcommand{\dvol}[1]{\der_{#1}\inn \vol}

\newcommand{\bd}{\mbox{\bf d}}
\newcommand{\bder}{\mbox{\bm $\der$}}
\newcommand{\bI}{\mbox{\bm $I$}}

\newcommand{\be}{\beta} 
\newcommand{\ga}{\gamma} 
\newcommand{\de}{\delta} 
\newcommand{\Ga}{\Gamma} 
\newcommand{\gmu}{\gamma^\mu}  
\newcommand{\gnu}{\gamma^\nu}
\newcommand{\ka}{\kappa}
\newcommand{\hka}{\hbar \kappa}
\newcommand{\al}{\alpha}
\newcommand{\lapl}{\bigtriangleup}
\newcommand{\psib}{\overline{\psi}}
\newcommand{\Psib}{\overline{\Psi}}
\newcommand{\derts}{\stackrel{\leftrightarrow}{\der}}
\newcommand{\what}[1]{\widehat{#1}}

\newcommand{\bx}{{\bf x}}
\newcommand{\bk}{{\bf k}}
\newcommand{\bq}{{\bf q}}

\newcommand{\omk}{\omega_{\bf k}} 

\newcommand{\lpl}{\ell}
\newcommand{\zb}{\overline{z}} 


\section{INTRODUCTION} 

In spite of the noticeable progress 
in the  quantum theory of gravity 
during the last decade, 
mainly owing to the 
Ashtekar program of non-pertubative canonical quantum 
gravity and the string/M--theory,  
it is 
difficult to escape a feeling that the 
genuine conceptual foundations of the 
synthesis of general relativity 
and the quantum theory are 
still awaiting of their discovery.   
In particular, the drastic differences between the way the physics 
is described general relativistically and quantum theoretically 
urge us to inquire if the presently known procedures of 
field quantization, 
that is   the specific way the quantum paradigm is implemented,  
are adequate to the problem of quantization of gravity. 
The 
particular concern is due to the different status of 
the time variable in quantum theory and in general relativity, 
in addition to the characteristic to the latter 
diffeomorphism covariance and a dynamical character 
of the space-time. 
The main objective of what we call the precanonical 
approach to field quantization is to elaborate a procedure 
which would treat all 
space-time variables on equal footing, more in accordance with 
the relativity theory. 
 
The idea of precanonical approach is suggested by 
a long-known in the calculus of variations 
 fact 
that the Hamiltonian 
formulation can be alternatively extended to field theory 
in the form of the 
De Donder-Weyl (DW) canonical equations \cite{dw}
\beq
\der_\mu y^a  = \frac{\der H}{\der p^\mu_a}, 
\quad 
\der_\mu  p^\mu_a=- \frac{\der H}{\der y^a }  ,  
\eeq 
 where given a Lagrangian density 
$L(y^a, y^a_\mu, x^\nu)$, a function of 
field variables $y^a$, their space-time derivatives 
(first jets) $y^a_\mu$ and   space-time 
variables $x^\mu$,  
one introduces new 
Ham\-il\-tonian-like variables  
 $p_a^\mu:=\der L / \der y^a_\mu$ ({\em polymomenta})  
 \mbox{\rm and} 
  $H=H(y^a,p^\mu_a,x^\nu):=\der_\mu y^a p^\mu_a -L$    
 ({\em the DW Hamiltonian function}). 
Obviously, this formulation 
is manifestly covariant, 
it treats space and time variables on equal footing, 
i.e. requires no usual $3+1$ decomposition,  
and corresponds to what could be viewed as a 
``multi-time'' generalization of 
the Hamiltonian formulation from mechanics to field theory. 
No picture of fields as infinite-dimensional mechanical systems evolving 
in time is implied in (1); 
instead, fields are described rather as systems varying in space and time. 
These features make 
formulation 
(1) an attractive alternative to the conventional ``instantaneous'' 
Hamiltonian formalism as a basis of quantization, 
especially in the context of general relativity.  
Besides, the obstacles to the 
DW Legendre transformation $y^a_\mu \rightarrow p_a^\mu$  in general 
are different from the usual constraints, 
thus 
suggesting a possibility 
of surmounting  the 
usual 
constraints analysis 
when quantizing. 

The term ``precanonical'' refers to the fact that 
 formulation 
(1) and the related 
constructions are in a sense  intermediate between the 
covariant Lagrangian description and the ``instantaneous'' canonical 
Hamiltonian description; in  mechanics precanonical structures 
coincide with the canonical ones while in field theory they 
are different.

Field quantization 
  stemming from  
formulation (1) has been considered 
in \cite{qs96,bial97,lodz98}; 
its application to general relativity 
has been discussed 
in a recent preprint by the author \cite{jena99}  
and will be outlined below. 
Briefly, quantization based on DW Hamiltonian formulation (1) 
leads to a generalization of quantum theoretic formalism in which 
the space-time Clifford algebra 
replaces 
the algebra of complex numbers (=the Clifford algebra of 
$(0+1)$--dimensional space-time!) in quantum mechanics. 
The Clifford algebra appears when quantizing 
the  Poisson brackets 
 which in DW theory are defined on differential forms \cite{romp98,bial96} 
(c. f.  \cite{regge}).     
This results in representing polymonenta by operators 
\beq
\hat{p}{}_a^\mu = -i\hbar\kappa \gamma^\mu \frac{\der}{\der y^a} , 
\eeq 
which act on spinor (or the Clifford algebra valued) 
wave functions $\Psi=\Psi(y^a,x^\mu)$; 
$\gamma^\mu${}'s denote the imaginary units of the space-time 
Clifford algebra;  the constant 
$\kappa$ of the dimension [length]$^{-3}$ ensures the dimensional 
consistency of (2) and is interpreted as a quantity of the ultra-violet 
cutoff or the fundamental length scale \cite{qs96,lodz98}. 
The wave function is supposed to fulfill a generalized Schr\"odinger 
equation 
\beq
i\hbar\kappa \gamma^\mu \der_\mu \Psi = \what{H}\Psi , 
\eeq
where $\what{H}$ is the operator form of the DW Hamiltonian function. 
This equation was found to be  consistent with several aspects of 
the correspondence principle \cite{bial97,lodz98}, 
for example, it leads to an analogue of 
the Ehrenfest theorem and 
can be reduced 
to the field theoretic  DW Hamilton-Jacobi equation  
(with some additional conditions) 
in the classical limit.  
 Unfortunately, the  details of the relationship 
 between 
 the standard quantum field  theory 
 and the present formulation 
 so far remain poorly understood. 
A possible connection with the Schr\"odinger functional 
picture has been  discussed in \cite{qs96,lodz98}. 


\section{PRECANONICAL QUANTUM GENERAL RELATIVITY: AN OUTLINE}

When applying the above approach to gravity 
the configuration space is to be  a bundle of symmetric 
second rank 
tensors over the space-time 
and the wave function is to be 
a function on this space: 
$\Psi = \Psi(g^{\mu\nu}, x^\mu)$.   
The Schr\"odinger equation for this 
wave function is obtained by first writing a curved space-time 
version of (3) and then replacing the metric and the connection 
by the corresponding operators. This leads to the following guess 
concerning the 
 wave equation for quantum gravity within the precanonical 
approach:  
\beq
i\hbar\kappa 
\what{ \mbox{$\hspace*{0.0em}e\hspace*{-0.3em}
\not\mbox{\hspace*{-0.2em}$\nabla$}$}} 
 \Psi = 
\what{{\cal H}\hspace*{-0.0em}} \Psi  ,   
\eeq 
 where   
$\what{{\cal H}}$ is the operator form 
of 
 DW Hamiltonian density  of gravity, 
${\cal H}:= \sqrt{g} H$, 
where $\sqrt{g} :=\sqrt {|{\rm det}(g_{\mu\nu})|}=:e$, 
and 
$\what{\mbox{$\hspace*{0.0em}\not\mbox{\hspace*{-0.2em}$\nabla$}$}} 
:= \gamma^\mu (\der_\mu + \hat{\theta}_\mu)$   
denotes the quantized covariant Dirac operator in which 
$\gamma$-matrices are such that 
$\gamma^\mu\gamma^\nu + \gamma^\nu\gamma^\mu= 2 g^{\mu\nu}$ 
and $\hat{\theta}_\mu$ is the spinor connection operator. 
Recall that classically 
$\theta_\mu = \theta^{\al\beta}{}_\mu \gamma_{[\al}\gamma_{\beta ]}$, 
 $\gamma^\mu = e^\mu_A \gamma^A$ 
 and 
\beq
\theta^{\al\beta}{}_\mu = 
g^{\nu [ \beta } ( 
\Gamma^{ \al ]}_{\mu\nu} -  e^{ \al ]}_A \der_\mu e^A_\nu 
) . 
\eeq  
Since the quantum gravity 
 possesses an intrinsic fundamental length scale, the Planck length 
$\lpl$, 
  one can expect that 
 $\kappa \sim \lpl^{-3}$.

 To find an operator realization of the quantities involved in 
(4) we first have to formulate the Einstein equations in 
DW Hamiltonian form. It is given by the set of equations 
\cite{horava} 
\beq  
\der_\al h^{\be\ga}
= 
\der {\cal H}  / \der Q^\al_{\be\ga} ,  
\quad 
\der_\al Q^\al_{\be\ga}
=  
- \der {\cal H}   / \der h^{\be\ga}  ,    
\eeq   
where 
$ h^{\alpha\beta}:= \sqrt{g}g^{\al\be}$ 
are field variables, 
\beq 
Q^\al_{\beta\gamma} := 
 \frac{1}{8\pi G} \left ( 
\delta^\alpha_{(\beta}\Gamma^\delta_{\gamma)\delta} 
  - \Gamma^\al_{\beta\gamma} \right ) 
\eeq  
are corresponding polymomenta 
and 
\beq
{\cal H} 
 :=  
 8\pi G\, 
h^{\alpha\ga} \left ( 
Q^\de_{\al\be } Q^\be_{\ga\de }
 - \frac{1}{3}\, Q^\be_{\al\be }Q^\de_{\ga\de } \right )
\eeq  
is the DW Hamiltonian function of gravity. 
 The above formulation of the Einstein equations 
 has a deeper foundation 
 in the theory of Lepagean equivalents in the calculus 
 of variations \cite{horava}. 

Now, polymomenta can be quantized according to the rule 
(2) adapted to curved space-time 
\beq
\what{Q}^\al_{\be\ga} = -i\hbar\kappa \gamma^\al 
\left \{  \sqrt{g} \frac{\der}{\der h^{\beta\gamma}} 
\right \}_{ord} , 
\eeq 
where the notation $\{ ... \}_{ord}$ refers to the 
ordering ambiguity of the expression inside the curly 
brackets. Plugging (9) into (8) we obtain 
\beq
\what{\cal H} = 
- \frac{16\pi}{3}  G \hbar^2 \kappa^2
 \left \{  \! \sqrt{g} h^{\al\gamma}h^{\beta\delta}
\frac{\der}{\der h^{\al\beta}}
\frac{\der}{\der h^{\gamma\delta}} \right \}_{\!ord}
\eeq

When formulating the left hand side of 
eq. (4) we are led to the 
fundamental difficulties 
related to the fact that (i) 
conceptually, the Dirac operator 
generally refers to a classical space-time 
background which is ought to be avoided in quantum gravity 
and (ii) technically, the last term in the 
spinor connection (5) cannot be expressed in 
terms of metric variables. 
 We deal with these difficulties by observing that 
the 
tetrads do not enter the present DW formulation 
of General Relativity  underlying the quantization 
and, therefore, can be treated only as 
non-quantized $x$-depended quantities $\tilde{e}^\mu_A(x)$. 
The correspondence 
principle then implies that they should be related to the 
mean value of the metric  
as follows     
\beq  
\tilde{e}{}^\mu_A(x)   
\tilde{e}{}^\nu_B(x) \eta^{AB} = 
\left < g^{\mu\nu}\right >(x)    
\eeq   
 where 
\beq
\left <g^{\mu\nu}\right >(x) = 
\int [d g^{\al\be}] 
~\Psib (g,x) g^{\mu\nu} \Psi(g,x)  ,   
\eeq 
and 
\beq
[d g^{\al\be}] = 
 {g}^{5/2} \prod_{\al \leq \be} d g^{\al\be}
\eeq   
is 
the invariant integration measure 
on the  $10$-dimensional space of metric components 
(c.f. \cite{misner}).  
Note that 
(12) is well-defined mathematically  
as a smooth field.

To quantize the connection coefficients let us note that classically 
(c.f. (7)) 
\beq 
\Gamma^\al_{\be\ga}= 8\pi G 
\left ( 
\frac{2}{3} 
\delta^\al_{(\beta } Q^\delta_{\ga)\delta} 
- Q^\al_{\be\ga} 
\right )  . 
\eeq 
Now, using (5) and (9) we can write 
\begin{eqnarray}{}
\what{\theta}^{\al\be}{}_{\mu} 
=   
-8\pi i G \hbar \kappa 
  \left \{ h^{\nu [\beta } 
\left ( 
\frac{2}{3} 
\delta^{ \al ]}_{(\mu } \gamma^\sigma 
\frac{\der}{\der h^{\nu)\sigma }} 
\right. \right. 
 \nn \\
 {}\quad - 
\left. \left. 
 \ga^{ \al ]} \frac{\der}{\der h^{\mu \nu}} 
\right ) \right \}_{ord} 
+\tilde{\theta}^{\al\be}{}_{\mu}(x) . 
\end{eqnarray} 
This expression involves the 
 ordering dependent 
operator  {} part { } $({\theta}^{\al\be}{}_{\mu})^{op}$  
and an  
 auxiliary spinor con\-nection part 
 $\tilde{\theta}^{\al\be}{}_{\mu}(x)$ which 
(i) accounts for the term in 
(5) which cannot be expressed in metric variables 
(hence, cannot be quantized) 
and 
(ii) ensures the transformation law 
of $\what{\theta}{}^{\al\be}{}_{\mu}$ 
is that of a spinor connection.  Our assumption is that 
$\tilde{\theta}{}^{\al\be}{}_{\mu}(x)$ is 
given by the standard formula  
 \beq
\tilde{\theta}^{\al\be}{}_\mu (x) = 
2 g^{\ga [\al} \tilde{e}{}^{\be ]}_B  
\der_{[\mu} \tilde{e}{}^{B}_{\ga ]}  
+ g^{\al\ga}g^{\delta\be}
\tilde{e}{}_\mu^B \der_{[\delta} \tilde{e}_{\gamma ] B}
\eeq 
where the tetrad field $\tilde{e}{}^\mu_A(x)$ 
 is given by (11). 

\newcommand{\spinconnection}{
\beq
\tilde{\theta}{}_\mu^{AB} (x) =  
\tilde{e}{}^{\al [A} 
\left 
( 
 2 \der_{[\mu} \tilde{e}{}^{B ]}{}_{\al ]} 
+ \tilde{e}{}^{ B] \be} \tilde{e}{}^C_\mu \der_\be 
\tilde{e}{}_{C \al}  
\right )
\eeq          } 

Now, we can formulate the 
diffeomorphism covariant wave equation for 
quantum  gravity: 
\beq
i\hbar\kappa 
\widetilde{ \mbox{$\hspace*{0.0em}e\hspace*{-0.3em}
\not\mbox{\hspace*{-0.2em}$\nabla$}$} 
} 
\Psi 
+  
i\hbar\kappa (\sqrt{g}  \gamma^\mu {\theta_\mu})^{op} \Psi 
=  \what{{\cal H}}\Psi   ,   
\eeq  
where 
$\widetilde{ \mbox{$\hspace*{0.0em} \hspace*{-0.3em}
\not\mbox{\hspace*{-0.2em}$\nabla$}$} } 
= \tilde{e}^\mu_A(x)\ga^A(\der_\mu + 
\tilde{\theta}_\mu (x))$ is the 
Dirac operator 
constructed using the self-consistent field 
$\tilde{e}^\mu_A(x)$ 
and  
\beq 
(\sqrt{g} \gamma^\mu {\theta_\mu})^{op} 
= - 4 
\pi i G \hbar\kappa 
\left \{ \sqrt{g} h^{\mu\nu} 
\frac{\der}{\der  h^{\mu\nu}} \right \}_{ord}  
\eeq  
is the term corresponding to the operator part 
of the spinor connection. The idea behind eq.~(17) 
is that classical geometric structures 
needed to formulate the wave equation are 
introduced as approximate averaged notions 
in a self-consistent with the underlying 
quantum dynamics 
(determined by $\what{{\cal H}}$) way.  
As a consequence of  condition (11) 
{} eq.~(17) essentially   becomes 
a non-linear  integro-differential equation 
describing the non-trivial way 
in which 
the wave function 
$\Psi$ specifies, or ``lays down'', 
the space-time geometry it propagates on.   

To complete the description, 
 we also need to impose a gauge-type 
condition in order to distinguish the physically relevant 
information. For example,  
the De Donder-Fock harmonic gauge  
can be  imposed in the form 
\beq 
\der_\mu \left <\sqrt{g}g^{\mu\nu}\right >(x) = 0  . 
\eeq 
In the present context this is a gauge condition 
on the wave function $\Psi(g^{\mu\nu}, x^\nu)$ 
rather than on 
the metric field.

\section{CONCLUSION }

The De Donder-Weyl Hamiltonian formulation of the field 
equations leads to the procedure of quantization of fields, 
which we suggested to call precanonical, 
 treating 
space and time variables on equal footing. When applied to general 
relativity in metric variables this framework  leads 
to a Dirac-like wave equation (17), 
non-linear and integro-differential,  
with self-consistently incorporated classical geometric 
structures.  No {\em arbitrarily} fixed background structure 
is present in the formulation; background independence 
is ensured by self-consistency which in its turn is dictated by the 
correspondence principle. The averaged self-consistent 
space-time serves its usual role: to order events 
(here, different possible configurations of the wave 
function in the metric space where its {\em linear} quantum 
dynamics is given by DW Hamiltonian operator 
$\what{\cal H}$) 
and to interpolate between them. This is the only way to 
describe physics we are sure about at the present:  
to describe it in space and time. 
Here we do have a quantum dynamics 
of the wave function in the metric space but we also do need 
classical space-time to order and join together the 
configurations  of the wave function in the metric space (the fibre) 
in different points of the space-time (the base). 
Our   wave equation prescribes how this ordering is achieved 
and how, as a consequence of this, the space-time is gaining 
its  metric structure. This reference to classical space-time, 
even though self-consistent, may well be an approximation.  
``Quantum space-time'', usually thought to be an essential 
ingredient of quantum  gravity, would imply a totally different 
way of describing physics. Our approach suggests that this 
could be achieved by attributing a proper sense to the 
``operator of the Dirac operator'' in the left hand side of (4) 
without referring to classical geometric notions. 
A. Connes' non-commutative geometry 
\cite{connes} can be mentioned  as an example of a mathematical 
framework achieving  this goal.

Potential advantages of the present approach are 
(i) the manifest covariance of the foundations and the results  
and (ii) the luck of serious mathematical problems with underlying 
mathematical constructions (as opposite to, e.g.,  
the Wheeler- De Witt geometrodynamics).  
It also offers a framework for discussing the problem of emergence 
of classical space-time in quantum gravity and 
has a potential to enlighten the problem of interpretation of quantum 
formalism in quantum cosmology: the quantum system described by 
(17) is  ``self-referential'' in the sense that 
the classical self-consistent tetrad field can be viewed as 
a model of the observing degrees of freedom explicitly entering 
into the description of quantum dynamics. We hope that these intriguing 
features are a sufficient  justification of the further 
analysis and development of the precanonical approach to 
quantum fields and quantum gravity, in spite of its so far 
unclarified  connections to the standard quantum field theory.   
\sloppy

\bigskip

\noindent 
{\bf Acknowledgments. }
\noindent 
I am grateful to the Organizers of the QG99  Conference 
for their kind invitation, warm hospitality, and the financial 
support which allowed  me to enjoy these unforgettable days in 
Villasimius (Sardinia).

\noindent  


\end{document}